# **Vorticity Flow Stabilization of Magnetized Plasmas**

F.Winterberg
University of Nevada
Reno, Nevada

Revised December 2009

# Abstract

Exact solutions of a magnetized plasma in a vorticity containing shear flow for constant temperature are presented. This is followed by the modification of these solutions by thermomagnetic currents in the presence of temperature gradients. It is shown that solutions which are unstable for a subsonic flow, are stable if the flow is supersonic.

The results are applied to the problem of vorticity shear flow stabilization of a linear z-pinch discharge.

*Instead of the word "shear flow" a better word is "vorticity flow", because not any flow with shear has vorticity, and it appears that only those flows which have vorticity can stabilize an unstable magnetohydrodynamic configuration.

# 1. Introduction

The linear z-pinch is the oldest and most simple magnetic plasma configuration, but it is unstable. The requirement for, stability can be relaxed if it is the goal to ignite a thermonuclear detonation wave propagation at supersonic velocities along the pinch discharge channel. There the time needed to keep the pinch stable is short [1, 2]. The ignition of the detonation wave could be done with a pulsed laser, for example. For the ignition of the thermonuclear detonation wave, the pinch current must be of the order $10^7$ Ampere to keep the fusion α-particles entrapped in the pinch, and the pinch must have a high density.

The possibility to stabilize a magnetized plasma by an axial shear flow has been considered by Hassam [3], and by Appl and Camenzind [4]. This was followed by three papers by Arber and Howell [5] by Pikuz et al. [6], and by Shumlak and Hartman [7]. The omission of Ohm's Law in the stability analysis by Arber and Howell, and also by Shumlak and Hartman, appears to be a serious defect. Arber and Howell only consider the magnetohydrodynamic equilibrium equation, which unlike Ohm's law is unchanged by the addition of an axial flow. And the paper by Shumlak and Hartman contains the following statement: "The presence of an axial flow does not affect the equilibrium equation." But the inclusion of Ohm's law for a plasma of infinite conductivity leads to a radial electric field, not taken into account in the stability analysis.

Apart from the quoted earlier papers, an extensive numerical simulation experiment was done and published by DeSouza-Machado, Hassam and Sina [8]. The paperwas followed by a detailed analytical study of the complete time dependent equations including the Hall term in Ohm's law by Sotnikov et al. [9]. A thorough stability analysis of the kink for different equilibrium profiles was done by Wanex, Sotnikov and Bauer [10], and by Wanex and Sotnikov [11] for both axial and radial shear flow, including magnetic shear. But it goes to the credit of Shumlak et al., [12, 13] to have actually carried out a large experiment. In this experiment a long co-axial plasma gun accelerates a supersonic annular plasma jet flowing over a z-pinch beginning from the end of the gun. However, the result of these experimrnts are not unique, because in one of them the pinch appears to be hollow [14].

## 2. Importance of vorticity and stagnation pressure

To appreciate the importance vorticity has for the stable confinement of a magnetized plasma it is worthwhile to compare the kink of current filament of a pinch discharge with a likewise kink of the vortex filament (see **Fig.1**). With **B** the magnetic field of the pinch discharge, one has outside the discharge channel curl **B** = 0, and likewise with **v** the velocity of a potential vortex, one has outside the vortex core curl **v** = 0. For a kink in a linear pinch discharge the magnetic field and hence the magnetic pressure is larger at the concave side of the kink, making the discharge unstable. For a line vortex, the velocity is likewise larger at the concave side of a kink, but because of Bernoulli's theorem this implies a lowering of the pressure at this side making a line vortex stable against a kink.

What is true for a line vortex with regard to the m=1 kink, is also true for the m=0 sausage instability. This follows from the conservation of the circulation,

$$Z = \oint \mathbf{v} \cdot d\mathbf{r} = \text{const.} \tag{1}$$

where the line integral is taken around the vortex filament. Because of (1) the velocity towards the inner surface of the vortex core rises as $1/r$, and $(1/2)\rho v^2$ as $1/r^2$. The magnetic pressure of a pinch discharge also rises at its surface as $1/r^2$. The magnetic pressure is trying to shrink the radius of the pinch, while the centrifugal force of the vortex flow is trying to do the opposite. This suggests to stabilize a pinch discharge by placing it inside a vortex. The discharge would there be stabilized if,

$$\left(\frac{1}{2}\right)\rho v^2 > \frac{B^2}{8\pi} \tag{1}$$

or if

$$v > v_A \tag{2}$$

where $v_A = B/\sqrt{4\pi\rho}$ in the Alfven velocity, implying that for stability $M_A>1$ is required, where $M_A = v/v_A$ is the Alfven Mach number. This qualitative conclusion is confirmed by the above cited experiment [12, 13].

While a current filament tends to get longer and break apart, a vortex filament has the opposite tendency. The reason for this very different behavior is the inertia of the fluid to which the vortex is attached, compared to the much smaller inertia of the magnetic field to which the pinch current is attached. The "stiffness" of a vortex filament, compared to "softness" of a current filament, is illustrated in **Fig.2**. The current density, $\mathbf{j} = c/4\pi$ curl $\mathbf{B}$, can be interpreted as the vorticity of the magnetic field, as $\boldsymbol{\omega} =$ curl $\mathbf{v}$ is the vorticity of fluid.

## 3. Vortex Sheet and Vortex Flow Configurations

Because it is difficult to place a pinch discharge inside a vortex, it has been proposed to superimpose upon the pinch discharge an axial, or more generally a spiraling flow. This idea though, has the effect that it removes heat from the pinch in the axial direction which can become a serious problem if the goal is to heat the pinch plasma to thermonuclear temperatures, but it is a welcome effect, if the goal is to launch a thermonuclear detonation wave moving supersonically down the pinch discharge channel requiring a pinch with the highest possible density, not a high temperature.

Instead of just using one vortex for the stable confinement of magnetized plasma, be it a linear pinch or any other plasma configuration, one may instead use a vortex flow made up of many vortices. The most simple of all vortex flow configuration is a vortex sheet [15]. It is established by the detachment of a flow along a boundary, as shown in **Fig.3**. A vortex flow is a smooth superimposition of many vortex sheets. One example of such a flow with constant vorticity is the Couette flow between two parallel plates. There one of the plates is at rest, while the other plate moves with a constant velocity against the plate at rest. A second example is a flow with uniform rotation, where curl **v** = const. throughout the entire fluid.

There are a variety of ways a vorticity flow, other than by the detachment from the boundary, can be realized. In the swirling Taylor flow for example, a uniform rotation is superimposed onto an axial flow [16]. It can be created by vanes placed in an axial flow. For a magnetized plasma rotation can be also generated by letting a plasma jet pass through a magnetic cusp. There the azimuthal component of the **j** × **B** force ($j_z$ × $B_r$) makes the plasma jet to rotate around the z-axis.

A nice example how the detachment from the boundary layer can create vortex sheet is the plasma focus (**Fig.4**). A circular vortex sheet is there created by detachment from the inner co-axial conductor, which in the plasma focus configuration collapses into a small volume, where it leads to the formation of many vortex filaments, explaining rather long life-time of the focus, which can be many Alfven times.

For the imposition of a vortex shear flow onto a linear z-pinch discharge one may inject a jet passing through the core of the pinch **Fig.5**, or inject an annular jet surrounding the pinch (see **Fig.6**). One way to generate a strong jet passing through the center of the pinch is by shooting a thin high velocity projectile through its core [17, 18] (see **Fig. 7**). If the magnetic pressure of the pinch discharge exceeds the tensile strength of the rod, the projectile acts like a dense plasma jet. This configuration has the advantage that the thin projectile can outside the pinch be accelerated to high velocities, by a light gas gun or a magnetic travelling accelerator, for example. As shown in **Fig.8**, a vorticity containing shear flow along a thin wire can also be obtained by the conical implosion of a plasma by a conical wire array onto a wire placed on the axis of the core. Another way to produce a high velocity jet is by the shape charge effect (see **Fig.9**), which can be realized by a conical wire array, with the wires imploded by a powerful electric discharge. However, in both instances there are large temperature gradients between the hot pinch plasma and the much cooler jet, leading to large thermomagnetic currents (Nernst effect), which must be included in the overall analysis.

One can also use an imploding wire array to implode a column of gas, which after it is compressed and heated, is injected as a jet along the pinch discharge channel. A better way still is to accelerate am magnetic plasma blob by a travelling magnetic wave to supersonic velocities which thereafter is injected into the pinch discharge (see **Fig.10**)

## 4. Exact Vortex Flow Solutions in Magnetohydrodynamics

The ideal magnetohydrodynamic equations for a plasma of infinite conductivity and zero viscosity are

$$\frac{\partial \mathbf{v}}{\partial t} = -\frac{1}{\rho}\text{grad } p - \frac{1}{2}\text{grad } \mathbf{v}^2 + \mathbf{v} \times \text{curl } \mathbf{v} - \frac{1}{4\pi\rho}\mathbf{B} \times \text{curl } \mathbf{B} \tag{3}$$

supplemented by

$$\frac{\partial \mathbf{B}}{\partial t} = \text{curl }(\mathbf{v} \times \mathbf{B}) \tag{4}$$

In a stationary solution of these quotations $\partial/\partial t = 0$. With no electric field inside the plasma this means that $\mathbf{v} \times \mathbf{B} = 0$, or that the stream lines $\mathbf{v}$ are aligned with the magnetic field lines $\mathbf{B}$.

In the absence of a vortex flow where $\text{curl } \mathbf{v} = 0$, the usual magnetohydrodynamic instabilities arise from the last term on the r.h.s of eq. (3). This means, these instabilities should be suppressed if

$$\mathbf{v} \times \text{curl } \mathbf{v} > \frac{1}{4\pi\rho}\mathbf{B} \times \text{curl } \mathbf{B} \tag{5}$$

If ρ=const, this means that

$$\mathbf{v} \times \text{curl } \mathbf{v} > \mathbf{v}_A \times \text{curl } \mathbf{v}_A \tag{6}$$

where, $\mathbf{v}_A = \mathbf{B}/\sqrt{4\pi\rho}$

Introducing the Alfven Mach number $M_A = v/v_A$, the inequality (6) then simply means that

$$M_A > 1 \tag{7}$$

Exact solutions for a stabilizing vortex flow can then be obtained from a solution of a differential equation

$$\mathbf{v} \times \text{curl } \mathbf{v} = \mathbf{a} \tag{8}$$

where **a** is a constant vector with $\mathbf{a} > \mathbf{v}_A \times \text{curl } \mathbf{v}_A$

For **a** = 0, eq. (8) describes a Beltrami flow, where the streamlines are aligned with the vortex lines. With **v** aligned to **B**, this is also the equation for force free magnetic field, where **B**×curl **B** = 0. Because there no magnetic forces act on the plasma, we exclude the case **a** = 0.

Setting

$$\mathbf{v} \times \text{curl } \mathbf{v} - \mathbf{v}_A \times \text{curl } \mathbf{v}_A = \mathbf{A} \tag{9}$$

where

$$\mathbf{A} = \left(1 - \frac{1}{M_A^2}\right) \mathbf{a} \tag{10}$$

for ∂/∂t = 0, and ρ = const. eq. (3) becomes

$$\operatorname{grad}\left(\frac{p}{\rho}+\frac{v^2}{2}\right)+\left(1-\frac{1}{M_A^2}\right)\mathbf{a}=0 \qquad (11)$$

Introducing a cylindrical r, $\varphi$, z coordinate system, for exact cylindrical symmetric solutions of (8) are obtained for $\partial/\partial\varphi = \partial/\partial z = 0$ and $v_r = 0$. The continuity equation div **v** = 0 is automatically satisfied. And the equation (8) has only a r-component.

We can therefore set

$$\mathbf{a} = \operatorname{grad}(ar) \qquad (12)$$

and hence for (11)

$$\operatorname{grad}\left[\frac{p}{\rho}+\frac{v^2}{2}-\left(1-\frac{1}{M_A^2}\right)ar\right]=0 \qquad (13)$$

which by integration yields

$$\frac{p}{\rho}+\frac{v^2}{2}-\left(1-\frac{1}{M_A^2}\right)ar = \text{const.} \qquad (14)$$

The last term in (14) can be viewed as the centrifugal potential of the vortex flow.

What remains is to solve (8) in cylindrical coordinates.

It there takes the form:

$$\frac{1}{2}\frac{dv_\varphi^2}{dr} + \frac{v_\varphi^2}{r} + \frac{1}{2}\frac{dv_z^2}{dr} = a \qquad (15)$$

Setting $v_\varphi^2 = y$, $r = x$ and $v_z^2 = z$,

this equation can be written as follows:

$$\frac{1}{2}\frac{dy}{dx} + \frac{y}{x} + \frac{1}{2}\frac{dz}{dx} = a \qquad (16)$$

This differential equation determines $v_\varphi(r)$ for a given $v_z(r)$ or vice verse $v_z(r)$ for a given $v_\varphi(r)$.

Let us consider a few simple cases:

I.     $v_z$ = const.

There $dz/dx = dv_z/dr = 0$, and hence

$$\frac{1}{2}\frac{dy}{dx} + \frac{y}{x} = a \qquad (17)$$

To solve this differential equation we set $y = xw$, hence

$$\frac{dy}{dx} = x\frac{dw}{dx} + w$$

and have

$$x\frac{dw}{dx} = 2a - 3w \tag{18}$$

with the solution (b constant of integration)

$$2a - 3w = (bx)^{-1/3} \tag{19}$$

or (different *b*)

$$y = \frac{x}{3}\left(2a - (bx)^{-1/3}\right) \tag{20}$$

which determines $v_\varphi = \sqrt{y}$ in terms of $v = x$.

    II.    We assume here that $v_z = bx$, hence

$v_z^2 = b^2 x^2$ , $(1/2)dv_z^2/dx = b^2 x$, and hence

$$\frac{1}{2}\frac{dy}{dx} + \frac{y}{x} + b^2 x = a \tag{21}$$

This differential equation must be numerically integrated.

III. Here we assume uniform rotation, setting $v_\varphi = br$, and have to determine $v_z = v_z(r)$

It is there more transparent if we go to eq. (15), which becomes

$$\frac{1}{2}\frac{dv_z^2}{dr} = a - 2b^2 r \qquad (22)$$

It has the solution (c constant of integration)

$$v_z = \sqrt{a(ar - b^2 r^2 + c)} \qquad (23)$$

In all these examples the stream lines and magnetic field lines spiral around the z-axis.

The solution obtained for a = const, is just one of a much larger number of possible solutions. Since it is obtained for $\mathbf{v} \times \mathrm{curl}\,\mathbf{v}$ = const. respectively $\mathbf{j} \times \mathbf{B} \propto \mathbf{v}_A \times \mathrm{curl}\,\mathbf{v}_A$ = const, it is the solution for a constant radial force density. If instead of the force density the current density shall be constant, then the force density is proportional to r. In this case, we have to replace a in eq. (11) by ar, and one has instead of eq. (14)

$$\frac{P}{\rho} + \frac{v^2}{2} - \frac{1}{2}\left(1 - \frac{1}{M_A^2}\right) ar^2 = \text{const.} \qquad (24)$$

and instead of (16)

$$\frac{1}{2}\frac{dy}{dx} + \frac{y}{x} + \frac{1}{2}\frac{dz}{dx} = ax \qquad (25)$$

Still more general solutions are obtained by replacing the constant a = |**a**| in eq. (11) by a = f(r), whereby (14) becomes

$$\frac{p}{\rho} + \frac{v^2}{2} - \left(1 - \frac{1}{M_A^2}\right)\int f(r)dr = \text{const.} \quad (26)$$

and instead of (16)

$$\frac{1}{2}\frac{dy}{dx} + \frac{y}{x} + \frac{1}{2}\frac{dz}{dx} = f(x) \quad (27)$$

With the help of eq. (15), eq. (26) can still be brought into different form, explicitly expressing the effect, the centrifugal force has on the solution.

From eq. (15) we have with a = f(r):

$$\frac{1}{2}\frac{dv^2}{dr} = \frac{1}{2}\left[\frac{dv_\varphi^2}{dr} + \frac{dv_z^2}{dr}\right] = f(r) - \frac{v_\varphi^2}{r} \quad (28)$$

Differentiating (26) with regard to r and expressing $(1/2)dv^2/dr$ by (28) we have

$$\frac{1}{\rho}\frac{dp}{dr} = \frac{v_\varphi^2}{r} - \frac{1}{M_A^2}f(r) \quad (27)$$

or

$$\frac{p}{\rho} = \int \frac{v_\varphi^2}{r}dr - \frac{1}{M_A^2}\int f(r)dr + \text{const.} \quad (28)$$

where

$$A = A(r) = \left(1 - \frac{1}{M_A^2}\right) f(r) \tag{29}$$

If the Mach number depends on r, whereby $M_A = M_A(r)$, eq. (26) is replaced by

$$\frac{p}{\rho} + \frac{v^2}{2} - \int \left(1 - \frac{1}{M_A^2}\right) f(r) dr = \text{const.} \tag{30}$$

and (28) by

$$\frac{p}{\rho} = \int \left[\frac{v_\varphi^2}{r} - \frac{1}{M_A^2(r)} f(r)\right] dr + \text{const.} \tag{31}$$

For the sake of transparency we have assumed that $\rho$ = const. We now generalize the solution for non-constant density, but in keeping with the assumption of cylindrical symmetry we still have $\partial/\partial\varphi = \partial/\partial z = 0$, and also $B_r = v_r = 0$. We start from eq. (3) for $\partial/\partial t = 0$:

$$\frac{1}{\rho} \text{grad} p + \frac{1}{2} \text{grad } v^2 - \mathbf{v} \times \text{curl}\mathbf{v} - \frac{1}{4\pi\rho} \mathbf{B} \times \text{curl}\mathbf{B} = 0 \tag{32}$$

and the equation of continuity

$$\text{div } \rho\mathbf{v} = \rho \text{ div } \mathbf{v} + \mathbf{v} \cdot \text{grad } \rho = \rho \text{ div } \mathbf{v} = 0 \tag{33}$$

because of $v_r = 0$. In (32) we have to replace $(1/\rho)$ grad p with grad w, where

$$w = \int \frac{dp}{\rho} \tag{34}$$

Next we transform the last term in (32) as follows:

$$\frac{1}{4\pi\rho}\mathbf{B}\operatorname{curl}\mathbf{B} = \frac{\mathbf{B}}{\sqrt{4\pi\rho}} \times \operatorname{curl}\left(\frac{\mathbf{B}}{\sqrt{4\pi\rho}}\right) + \frac{\mathbf{B}}{\sqrt{4\pi\rho}} \times \mathbf{B} \times \operatorname{grad}\left(\frac{1}{\sqrt{4\pi\rho}}\right)$$

$$= \mathbf{v}_A \times \operatorname{curl}\mathbf{v}_A + \mathbf{B}\left(\frac{\mathbf{B}}{\sqrt{4\pi\rho}} \cdot \operatorname{grad}\frac{1}{\sqrt{4\pi\rho}}\right) - \frac{B^2}{\sqrt{4\pi\rho}}\operatorname{grad}\left(\frac{1}{\sqrt{4\pi\rho}}\right)$$

$$= \mathbf{v}_A \times \operatorname{curl}\mathbf{v}_A + \frac{B^2}{8\pi\rho^2}\operatorname{grad}\rho \tag{35}$$

because of $B_r = 0$.

With this result we go back into (32) and have

$$\operatorname{grad} w + \frac{1}{2}\operatorname{grad} v^2 + \frac{B^2}{8\pi\rho^2}\operatorname{grad}\rho + \mathbf{v}\times\operatorname{curl}\mathbf{v} - \mathbf{v}_A\times\operatorname{curl}\mathbf{v}_A = 0 \tag{36}$$

or

$$\operatorname{grad} w + \frac{1}{2}\operatorname{grad} v^2 + \frac{B^2}{8\pi\rho^2}\operatorname{grad}\rho + \mathbf{A} = 0 \tag{37}$$

Of special interest is the case where $B/\rho$ = const. It is called the "frozen in" condition expressed by the Walen equation [19, 20]. It has the "equation of state" for a gyrotropic plasma where the plasma density is inversely proportional to the square of the Larmor radius for the plasma ions of mass M:

$$r_L = \frac{Mv_L c}{ZeB} \tag{38}$$

where $v_L$ is thermal ion velocity perpendicular to **B**.

There then,

$$\frac{B^2}{8\pi\rho^2} \operatorname{grad} \rho = \frac{1}{2} \operatorname{grad} v_A^2 \tag{39}$$

whereby (37) becomes

$$\operatorname{grad} w + \frac{1}{2}\operatorname{grad}(v^2 + v_A^2) + \mathbf{A} = 0 \tag{40}$$

hence

$$w + \frac{1}{2}(v^2 + v_A^2) + \int A(r)dr = \text{const.} \tag{41}$$

If B/ρ is not constant, we have instead

$$w + \frac{1}{2}v^2 + \int \left[\frac{1}{2}v_A^2(r)\frac{d \ln \rho}{dr} + A(r)\right] dr = \text{const.} \tag{42}$$

## 5. Stability

Stability can most generally determined by energy principle of Bernstein, Frieman, Kruskal and Kulsrud [21]. It is most easily established for constant density $\rho$ and if $M_A>1$. For plasmas where $\beta = 1$, ($\beta=nkT/(B^2/8\pi)$), $M_A>1$ means that the flow is supersonic. If $\rho$=const., (1) becomes

$$\frac{\partial \mathbf{v}}{\partial t} = -\nabla\left(\frac{p}{\rho} + \frac{v^2}{2}\right) + \mathbf{v} \times \text{curl}\mathbf{v} - \frac{1}{4\pi\rho}\mathbf{B} \times \text{curl}\mathbf{B} \tag{43}$$

Or with $\mathbf{v}\|\mathbf{B}$, $v/v_A = M_A$

$$\frac{\partial v}{\partial t} = -\nabla\left(\frac{p}{\rho} + \frac{v^2}{2}\right) - (1 - M_A^2)\frac{1}{4\pi\rho}\mathbf{B} \times \text{curl}\mathbf{B} \tag{44}$$

with (2) remaining unchanged.
Setting

$$p \rightarrow p/\rho + v^2/2$$
$$\mathbf{B} \rightarrow \sqrt{1 - M_A^2}\, \mathbf{B} \tag{45}$$

and setting div v=0 for $\rho$=const., one obtains from the energy principle that

$$\delta W = \frac{1}{4\pi}\int (1 - M_A^2)[Q^2 - (\nabla \times \mathbf{B}) \cdot (\mathbf{Q} \cdot \boldsymbol{\epsilon})]d^3x > 0 \tag{46}$$

where

$$\mathbf{Q} = \nabla \times (\boldsymbol{\xi} \times \mathbf{B})$$
$$\boldsymbol{\xi} = \int \mathbf{v}\, dt \tag{47}$$

Therefore, configurations which for $M_A<1$ are unstable, are stable for $M_A>1$, that means for $\beta=,1$they are stable for supersonic flow. This conclusion is confirmed in at least one experiment done by Shumlak et al. [12, 13]. It showed that the otherwise unstable linear z-pinch, can be stabilized by a supersonic flow in the axial direction.

The condition $M_A>1$ means simply that

$$\rho v^2 > B^2/4\pi \qquad (48)$$

If $B^2/4\pi \gg \rho v^2$, the magnetic pressure overwhelms the fluid stagnation pressure, and it "carriers along" the fluid, while for $\rho v^2 \gg B^2/4\pi$ the opposite is true, with the flow carrying along the magnetic field [20, 21].

If the plasma is assumed to be compressible, additional terms have to be added to the integrand of (46). They are all positive and do not lead to instabilities.

An entirely different situation arises if the flow becomes turbulent, or if through discontinuities in the plasma, either inside the plasma or by the boundary, shock waves occur, if $M_A>1$. In general, the plasma flow becomes turbulent above a critical Reynolds number. For magnetic plasma confinement turbulence is highly undesirable because it can drastically reduces the energy confinement time.

## 6. The Effect of a Temperature Gradient on a Vortex Sheet

The foregoing analysis was restricted to plasma of constant density and constant temperature. If there is a temperature gradient across the vortex sheet it induces in the presence of a magnetic field a thermomagnetic current (Nernst effect), modifying the flow and magnetic field. A temperature gradient across a vortex sheet exists even if the two flows separated from each other have the same temperature, because the stagnation of the two fluids coming to a relative rest leads to a difference in their temperature.

What is true for two flows separated by a vortex sheet, is also true for a flow along a cold wall. There too, the presence of a magnetic field is going to induce thermomagnetic currents, changing the magnetic field and flow.

In the presence of a temperature gradient $\nabla T$ and a magnetic field **B**, the density of the thermomagnetic Nernst current $\mathbf{j}_N$ is given by [24]

$$\mathbf{j}_N = \frac{3nkc}{2B^2} \mathbf{B} \times \nabla T \tag{49}$$

It exerts a force on plasma, with the force density given by

$$\mathbf{f} = \frac{1}{c} \mathbf{j}_N \times \mathbf{B} = \frac{3\,nk}{2\,B^2} (\mathbf{B} \times \nabla T) \times \mathbf{B} \tag{50}$$

With $\nabla T$ perpendicular to **B** one has

$$\mathbf{f} = \frac{3}{2} nk \nabla T \tag{51}$$

But since also

$$\mathbf{f} = \nabla p \tag{52}$$

And because $p = 2nkT$, for a fully ionized hydrogen plasma, one has

$$\nabla p = 2nk\nabla T + 2kT\nabla n \tag{53}$$

and because of (45) and (46)

$$n\nabla T + 4T\nabla n = 0 \tag{54}$$

In the important case of plasma flow along a solid wall of zero temperature, the integration of (48) yields

$$Tn^4 = T_o n_o^4 \tag{55}$$

Where $T_o$, $n_o$ are the stagnation temperature and particle number density, of the plasma coming to rest at the wall.

The stagnation temperature is given by [25]

$$T_o = T\left(1 + \sqrt{P}\,\frac{\gamma-1}{2}\,M_o^2\right) \tag{56}$$

Where $M_o > 1$ is the Mach number of the plasma flow, and **P** the Prandtl number. For a fully ionized hydrogen plasma one has $\mathbf{P} = \gamma = 5/3$, where $\gamma$ is the specific heat ratio $c_p/c_v$.

Inserting the expression of the Nernst current into the Maxwell equation

$$\frac{4\pi}{c}\mathbf{j}_N = \text{curl }\mathbf{B} \tag{57}$$

one obtains

$$6\pi n k \mathbf{B} \times \nabla T = B^2\,\text{curl }\mathbf{B} \tag{58}$$

With **B** directed along the wall it becomes

$$6\pi n k \frac{dT}{dr} = -B\frac{dB}{dr} \tag{59}$$

or because of (55)

$$B\,dB = -6\pi k n_o T_o^{1/4} T^{-1/4}\,dT \tag{60}$$

Integrating (60) through the boundary from $B = B_o$, $T = T_o$ at the wall into the plasma, one finds that

$$\frac{B_o^2}{8\pi} = 2n_o kT_o \qquad (61)$$

This means that the magnetic field of the thermomagnetic Nernst current insulates the hot plasma against the wall.

With $Tn^4$=const, the bremsstrahlungs losses near the wall are substantially reduced. These losses are given by

$$\varepsilon_r = 1.42 \times 10^{-27} n^2 \sqrt{T} \quad \text{erg/cm}^3 \text{s} \qquad (62)$$

without the Nernst current where $Tn$ = const, $\varepsilon_r$ would rise as $T^{-3/2}$, with $T \to 0$, in approaching the wall, while for $Tn^4$ these losses remain constant.

# 7. Discussion

The idea to stabilize a dense z-pinch by shooting at high velocities a thin metallic projectile through its core remained unnoticed for many years until the appearance of paper by Hassam and Huang [26], who proposed to inject a high velocity plasma jet into a large toroidal tube. Apart from the stabilization of the flowing plasma through vorticity generating shear by its interaction with the wall of the toroidal tube, both concepts have in common the importance of the Nernst effect. It results from the large temperature gradient between the cool projectile and the hot pinch plasma, respectively between the cool wall and the hot plasma jet.

In a number of papers which have treated the problem of stabilizing the linear z-pinch discharge with a superimposed shear flow it has been assumed that the undisturbed configuration has a superimposed axial flow, but no axial magnetic field, and no azimuthal flow. This assumption is very problematic, because in the presence of a strong azimuthal magnetic field, the superimposed axial flow is deformed obtaining an azimuthal component, and the azimuthal magnetic is likewise deformed obtaining an axial component. In other words, the equilibrium configuration must consist of helically aligned magnetic field- and stream-lines. The above made assumption is only justified for a small axial flow, perturbing the static equilibrium, but not for large disturbances at large Mach numbers. Perturbation theory for the study of stability is only possible for a non-perturbatively obtained solution, or at best for a slightly perturbed equilibrium at small Mach numbers of the superimposed axial flow.

Because of the stiffness of a vorticity containing flow, the obtained exact solutions should be stable at high Mach numbers. Only at very large Mach numbers, with large Reynolds numbers can the flow become unstable through the onset of turbulence. But it is unlikely that the Mach numbers sufficient to achieve stability, will surpass this limit.

It is also interesting to know the direction of the stiffness generating vorticity vector. For an axial shear flow the vorticity is directed azimuthally with $\mathrm{curl}_\varphi \mathbf{v} = dv_z/dr$. For an azimuthal flow it goes into the axial direction. No twisting of the magnetic field and stream lines only happens for an azimuthal flow in a linear z-pinch, and an axial flow in the linear theta pinch.

Very much as an otherwise unstable magnetized plasma can be wall-stabilized by a conducting wall, one may likewise stabilize a magnetized plasma by surrounding it with a non-magnetized vorticity containing plasma. In this way one can even obtain stable solutions where a magnetized plasma sphere is confined inside a vorticity containing spiral flow [27].

Finally, we could like to compare the configuration **Fig. 7** of a thin metallic rod (i.e. wire) shot through the center of a pinch discharge, with the configurations shown in **Fig. 8,** where an imploding conical wire array produces a jet along a thin rod over which a large current is flowing. In the rod through the pinch configuration, the pinch is attached to the large mass of the earth through the magnetic field of the pinch discharge- producing apparatus bolted to the ground, while in the conical implosion onto a central rod it is the rod which is attached to the earth. In either case we have to deal with a two-body problem, but because the mass of the earth is very large, this two body problem must in either case be analyzed in the rest frame of the earth. The crucial number in each experiment is the momentum flux density relative to the earth that is relative to the laboratory. For a rod with a density of $\rho \sim 10 \text{g/cm}^3$, and a velocity of $10 \text{km/s} = 10^6 \text{cm/s}$, the momentum flux density is $\rho v^2 = 10^{13} \text{dyn/cm}^2$. For the plasma accelerated into an axial direction by a conical wire implosion "shape charge", the plasma density is of the order $\rho \simeq 10^{-3}$ g/cm, moving with a velocity of $100 \text{km/s} = 10^7 \text{cm/s}$, and one there has $\rho v^2 = 10^{11} \text{dyn/cm}^2$, is about 100 times smaller. For the 10km/s moving rod, the momentum flux density $\rho v^2 = 10^{13} \text{dyn/cm}^2$, is about $10^3$ times larger than the tensile strength of the rod, which is why the moving rod can be viewed as a dense plasma jet, able to overcome a magnetic pressure $B^2/8\pi = \rho v^2$, or a magnetic field of about $10^7$ Gauss. By comparison, with a 100 times smaller momentum flux density, at $\rho \sim 10^{-3} \text{g/cm}^3$ and $v \sim 100 \text{km/s}$, the accelerated plasma can only overcome a magnetic field which is ten times less.

Another way to achieve the same as for a rapidly moving rod, is to make use of the shape charge effect as shown in **Fig.9**, where for metallic liners jet velocities of ~10km/s are possible [14]. There the high velocity metallic jet from the shape charge assumes the role of the rod shot through a pinch. This concept though has the drawback that a shape-charge produced jet cannot be easily collimated to have the same small diameter as a needle-like projectile, accelerated to the same velocity.

Since the rapidly moving rod can be viewed as a cool plasma moving at ~ 10km/s, and since the velocity of sound in this plasma is by order of magnitude equal to $\sqrt{\sigma/\rho}$, where $\sigma \sim 10^{11} \text{dyn/cm}^2$, is the tensile strength of the rod, the velocity of sound in the rod is about 1km/s, about equal the Alfven velocity, making $M_A \sim 10$. To reach such a large Alfven Mach number is much more difficult if the shear flow stabilization is done by a plasma jet. For a z-pinch at thermonuclear temperatures the Alfven velocity is about $10^8$ cm/s. To reach a stabilizing plasma flow at $M_A \sim 10$, would require to heat a plasma to $10^{10}$ °K, or to accelerate it by travelling magnetic wave to a velocity of $\sim 10^9$ cm/s, as shown in **Fig.10**. Instead of using a travelling magnetic wave accelerator, a co-axial plasma gun was proposed for the same purpose by Shumlak et al. [12], but there not the kind of very large Mach numbers can be reached.

## 6. Conclusion

This study presents a number of exact stationary vortex flow solutions in classical magnetohydrodynamics, which unlike similar static solutions are expected to be stable. For the attainment of this stability through, a price has to be paid. It is the energy loss of the confined plasma by the vorticity containing flow, but this energy loss is unimportant for the fast thermonuclear detonation wave ignition of cool, but dense, magnetically confined plasma.

The analysis suggests that the stability is improved by increasing the Alfven Mach number. The flow however, must be kept below the critical Reynolds number above which turbulence sets in. This limits the Mach number to a value which still has to be determined, both theoretically and experimentally.

The assumption that $\partial/\partial t = 0$ leaves out an important aspect: It is the formation of shock waves. For the annular injection of a supersonic plasma jet it results in (to the pinch axis directed) convergent cylindrical shock waves, which can substantially contribute to the heating of the pinch.

**Figures**

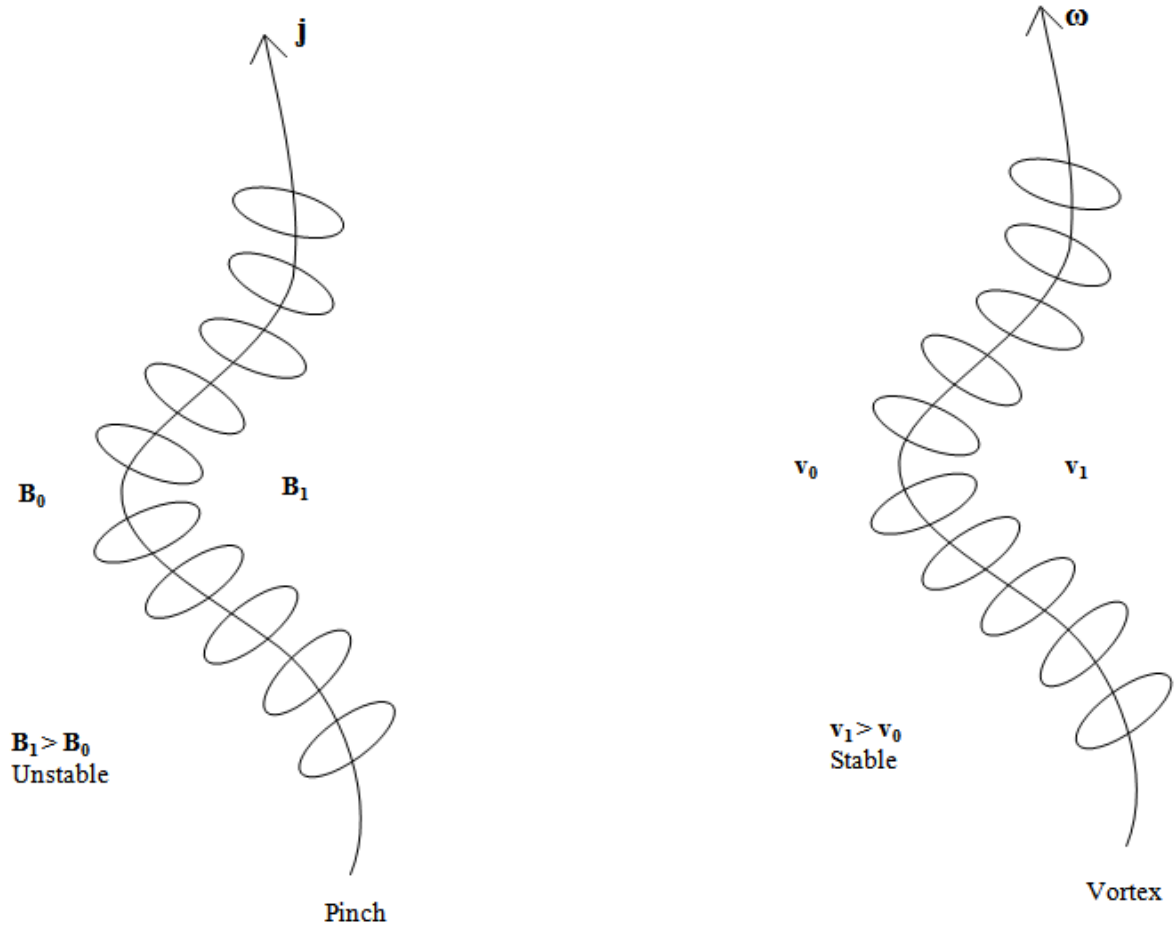

**Fig.1** Comparison of the current filament with a line vortex filament.

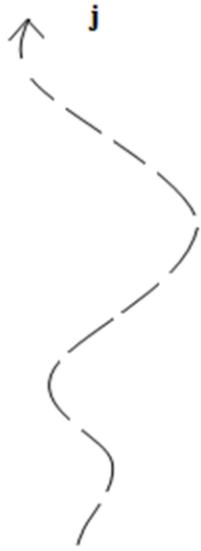

Current filament  Vortex filament
(soft, with breakup)  (stiff)

**Fig.2** The unstable current filament versus the stable vortex filament.

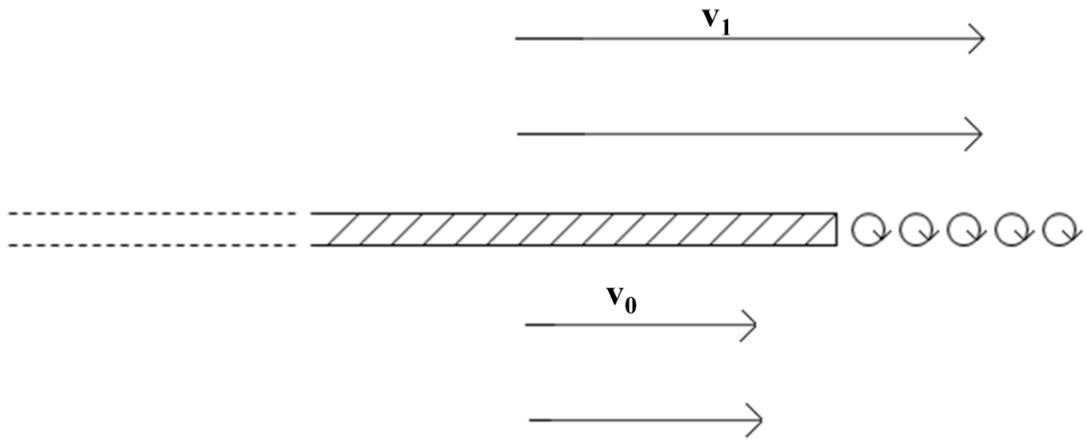

**Fig.3** Formation of vortex sheet, $v_1 > v_0$.

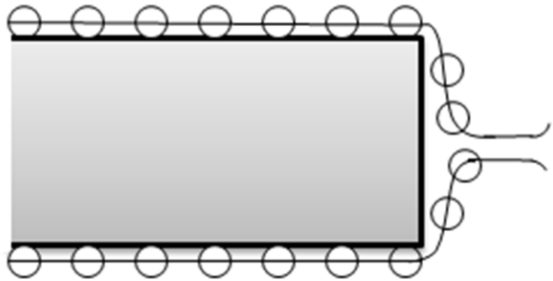

**Fig.4** Plasma Focus: An example of a flow with azimuthal vorticity.

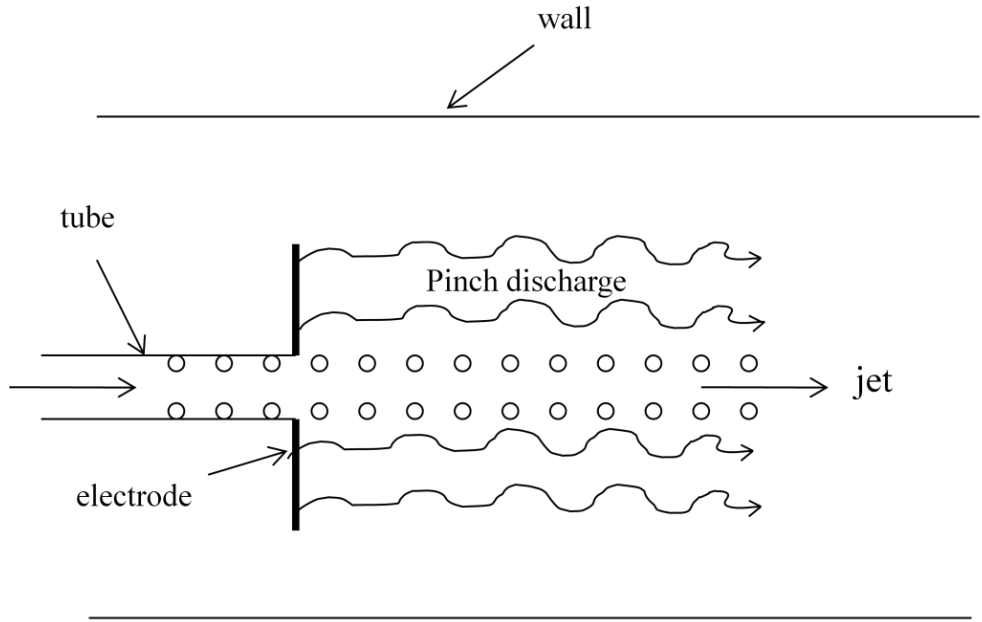

**Fig. 5** Shear flow (vorticity) stabilization of linear z-pinch discharge with central jet.

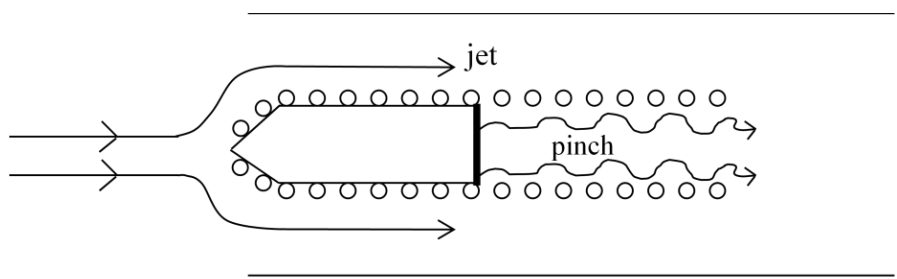

**Fig.6** Shear flow (vorticity) stabilization of linear z-pinch discharge with annular jet.

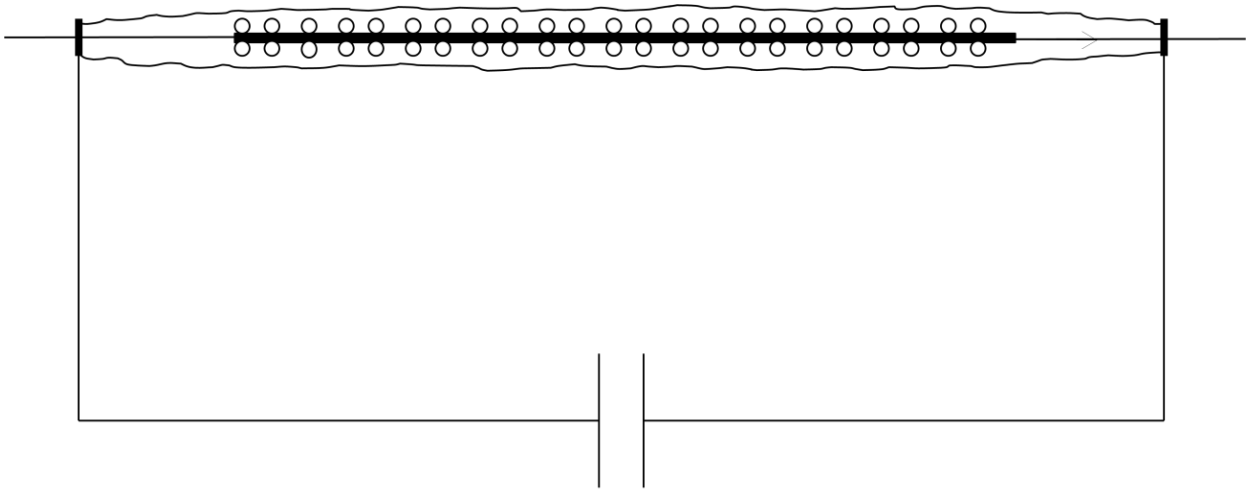

**Fig. 7** Shear flow vorticity stabilization with high velocity injection of a metallic jet through the core of a pinch discharge.

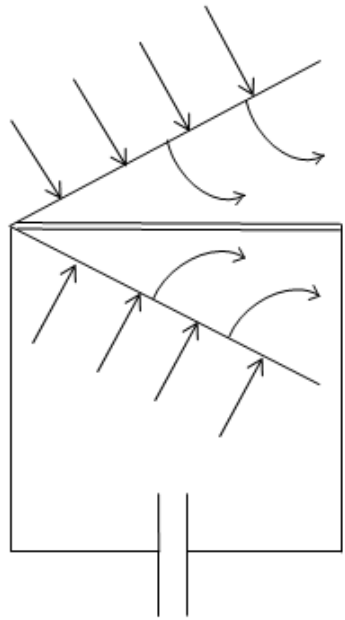

**Fig.8** Conical implosion of plasma onto a central rod over which a pinch discharge takes place.

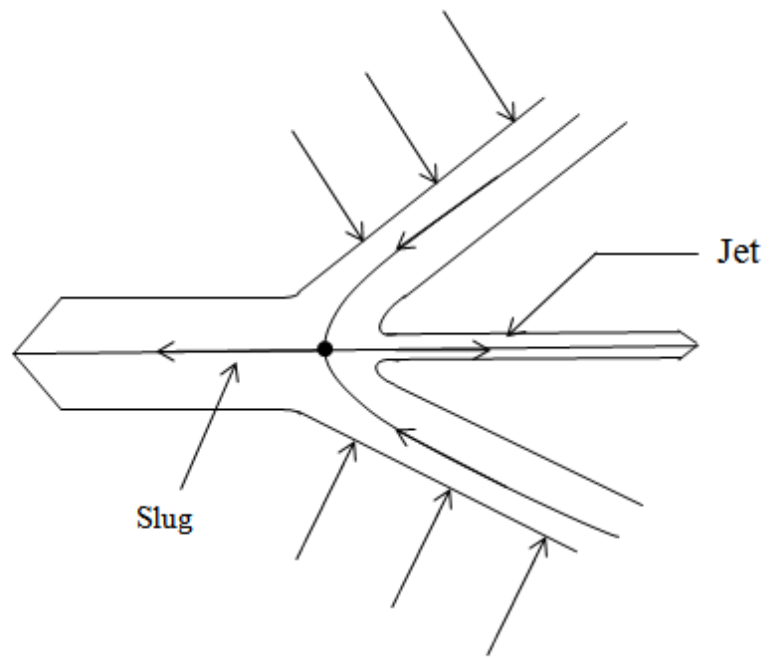

**Fig.9** Jet formation in a conical implosion.

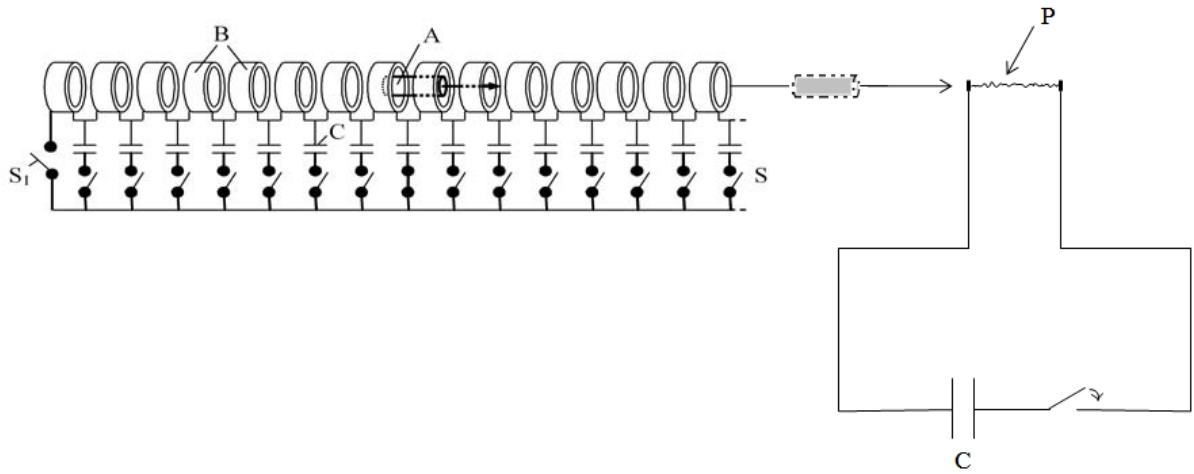

**Fig.10** Hypervelocity acceleration and injection of magnetized plasma cylinder onto a linear pinch discharge. A magnetized plasma blob; B magnetic field coil; C capacitor banks; S and S1 switches; P pinch discharge.